\documentclass[epj]{svjour}
\usepackage{graphicx}
\usepackage{amsmath}
\usepackage{amsfonts}
\usepackage{amssymb}
\usepackage{epsfig}
\usepackage{times}
\usepackage{calc}
\usepackage{version}
\usepackage[french,english]{babel}

\def\kt{{k_{_\perp}}}
\def\cO#1{{{\cal{O}}}\left(#1\right)}

\newcommand{\lqcd}{\Lambda_{_{_{\rm QCD}}}}
\newcommand{\dd}{{\rm d} }

\begin{document}
\title{Transverse momentum spectra of charged hadrons in jets at the
Tevatron}
\author{Redamy Perez Ramos \inst{1}
\thanks{\emph{Luruper Chaussee 149, D-22761 Hamburg, Germany}}%
}                     
\offprints{Redamy Perez Ramos}          
\institute{II. Institut f\"ur Theoretische Physik, 
Universit\"at Hamburg}
\date{Received: date / Revised version: date}
%
\abstract{The hadronic $\kt$-spectrum inside a high energy
 jet is determined including corrections of relative magnitude
 $\cO{\sqrt{\alpha_s}}$ with respect to the Modified Leading Logarithmic
Approximation (MLLA),  in the limiting spectrum approximation
(assuming an infrared cut-off $Q_0 =\lqcd$) and beyond ($Q_0\ne\lqcd$).
The  results in the limiting spectrum approximation
are found to be, after normalization, in
impressive agreement with measurements by the CDF
collaboration. A new integral representation for the
``hump-backed" plateau is also reported.
} 
\maketitle
\section{Introduction}
\label{intro}
The production of jets --~a collimated bunch of hadrons~--
in $e^+e^-$, $e^-p$ and hadronic collisions is an ideal playground
to investigate the parton evolution process in perturbative QCD (pQCD).
One of the great successes of pQCD is the existence of the hump-backed
shape of inclusive spectra, predicted in~\cite{HBP} within MLLA,
and later discovered experimentally~(for review, see e.g.~\cite{KhozeOchs}).
Refining the comparison of pQCD calculations
with jet data taken at LEP, Tevatron and LHC will ultimately allow for
a crucial test of the Local Parton Hadron Duality (LPHD)
hypothesis~\cite{LPHD} and for a better understanding of color
neutralization processes. 

The inclusive $\kt$-distribution of particles
inside a jet has been computed at MLLA accuracy
in the limiting spectrum
approximation~\cite{PerezMachet}, {\it i.e.} assuming an infrared cutoff $Q_0$ equal to
 $\lqcd$ ($\lambda\equiv\ln( Q_0/\lqcd)= 0$) (for a review, see also \cite{Basics}).
MLLA corrections, of relative magnitude $\cO{\sqrt{\alpha_s}}$ with respect
to the leading double logarithmic approximation (DLA), were shown to be quite
substantial ~\cite{PerezMachet}. Therefore, it appears legitimate
to wonder whether corrections of order $\cO{\alpha_s}$, that is
next-to-next-to-leading or Next-to-MLLA (NMLLA), are negligible or not.

The starting point of this analysis is the MLLA Master Equation for
the {\em Generating Functional} (GF) $Z=Z(u)$
of QCD jets \cite{Basics}, where $u=u(k)$ is a certain probing
function and $k$, the four-momentum of the outgoing parton.
Together with the initial condition at threshold, the GF determines the
jet properties at all energies. For instance, the single inclusive 
spectrum can be derived from
the GF by differentiating with respect to $u=u(k)$,
and the solution of the equations can be written as a perturbative
expansion in $\sqrt\alpha_s$.
At high energies this expansion can be resummed and the leading contribution
be represented as an exponential of the anomalous dimension 
$\gamma(\alpha_s)$. Since further details to this logic
can be found in \cite{Basics,RPR2}, we only
give the symbolic structure of the equation for the GF 
and its solution, which we write respectively as
\begin{equation}\label{eq:symbGF}
\frac{dZ}{dy}\simeq\gamma_0(y)Z\quad\Rightarrow\quad
Z\simeq\exp{\left\{\int^y\gamma(\alpha_s(y'))dy'\right\}},
\end{equation}  
where $\gamma(\alpha_s)$ can be expressed as a
power series of $\sqrt\alpha_s$
\begin{equation}\label{eq:anomdim}
\gamma(\alpha_s)=\,\sqrt\alpha_s+\,\alpha_s+\,\alpha_s^{3/2}
+\,\alpha_s^2+\ldots
\end{equation}
In this logic, the leading 
(DLA, ${\cal O}(\sqrt{\alpha_s})$) 
and next-to-leading (MLLA, ${\cal O}(\alpha_s)$) approximations are complete.
The next terms (NMLLA, ${\cal O}(\alpha_s^{3/2})$) 
are not complete but they include an important
contribution which takes into account energy conservation and an
improved behavior near threshold. Indeed, some results for such NMLLA terms have been
studied previously for global observables and have been found to
better account for recoil effects and to drastically affect
multi-particle production \cite{DokKNO,CuypersTesima}. 

The equation in (\ref{eq:symbGF}) applies to each vertex
of the cascade and its solution represents the fact that 
successive and independent partonic 
splittings inside the shower,
exponentiate with respect to the {\em evolution-time} parameter 
$dy=d\Theta/\Theta$, where $\Theta\ll1$ is the angle
between outgoing couples of partons. The choice of $y$
follows from Angular Ordering (AO) in intra-jet cascades.
It is indeed the suited variable for describing {\em time-like}
evolution in jets. Thus, Eq.~(\ref{eq:symbGF}) incorporates the Markov chains
of sequential angular ordered partonic decays which are singular in $\Theta$
and $\gamma(\alpha_s)$ determines the real rate of inclusive quantities
growth with energy.

While DLA treats the emission of both particles as independent
by keeping track of the first term $\sim\sqrt\alpha_s$
in (\ref{eq:anomdim}) without constraint,
the exact solution of the MLLA evolution equation (partially) fulfills
the energy 
conservation in each individual splitting process ($z+(1-z)=1$) by
incorporating higher order ($\alpha_s^{n/2},n>1$) terms to the anomalous 
dimension. Symbolically, the first two analytical steps towards a better
account of these corrections in the MLLA, NMLLA evolution,
which we further discuss in \cite{PAM}, 
can be represented in the form
\begin{equation}\label{eq:deltagamma}
\Delta\gamma\simeq\int(\alpha_s+\alpha_s\ell^{-1}\ln z)dz\sim\alpha_s+
\alpha_s^{3/2},
\end{equation}
where $\ell=\ln(1/x)\sim\alpha_s^{-1/2}$ with $x\ll1$ (fraction of the
jet energy taken away by one hadron), $z\sim1$ for
hard partons splittings such as $g\to q\bar q$\ldots (this is
in fact the region where the two partons are strongly
correlated). 

Energy conservation is particularly important for energetic particles
as the remaining phase space is then very limited.
On the other hand, a soft particle can be emitted with little impact on 
energy conservation. Some consequences of this behavior have also been
noted in \cite{KLO}:

\begin{itemize}
\item[$(i)$] the soft particles follow the features expected from DLA;
\item[$(ii)$] there is no energy dependence of the soft spectrum;
\item[$(iii)$] the ratio of soft particles $r=N_g/N_q$
in gluon and quark jets is consistent with the DLA prediction $N_c/C_F=9/4$
(see the measurement by DELPHI \cite{lepdelphi}).
\end{itemize}

The present study makes use of this logic (\ref{eq:deltagamma}) to evaluate NMLLA
contributions to the single
inclusive $k_\perp$-distribution. The main results
of this work have been published in \cite{PAM,PRL}.
Experimentally, the CDF collaboration at the Tevatron reported on
$\kt$-distributions of unidentified charged hadrons in jets produced in
$p\bar{p}$ collisions at $\sqrt{s}=1.96$~TeV~\cite{CDF}.
\section{MLLA evolution equations}
\label{section:sis}
\begin{figure}[ht]
\begin{center}
  \includegraphics[height=2.2cm]{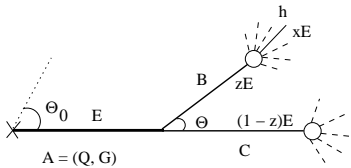}
\end{center}
\caption{Parton A with energy $E$ splits into parton B
(respectively C) with energy $zE$ (respectively, $(1-z)E$)
which fragments into a hadron $h$ with energy $xE$.}
\label{fig:spplit}
\end{figure}

We start by writing the MLLA evolution equations for the fragmentation
function $D_{B}^h\left(x\big/z, zE\Theta_0, Q_0\right)$ of
a parton B (energy $zE$ and transverse momentum $\kt=zE\Theta_0$)
into a gluon (represented by a hadron $h$ (energy $xE$)
according to LPHD \cite{LPHD}) 
inside a jet A of energy $E$ for the process depicted
in Fig.~\ref{fig:spplit}.
As a consequence of angular ordering in parton cascading,
partonic distributions inside a quark and gluon jet,
$Q,G(z)=x\big/z\ D_{Q,G}^h\left(x\big/z, zE\Theta_0, Q_0\right)$, obey
the system of two coupled  equations \cite{RPR2} (the subscript ${}_y$ denotes
$\partial/\partial{y}$) following from the MLLA master equation
described by (\ref{eq:symbGF})
\begin{eqnarray}
 Q_{y}&=&\!\!\int_0^1 \dd z\>  \frac{\alpha_s}{\pi} \>\Phi_q^g(z)\>
\bigg[ \Big( Q(1-z)-Q\Big) +  G(z) \bigg],\label{eq:qpr}\\
 G_{y}&=&\!\!\int_0^1 \dd z\> \frac{\alpha_s}{\pi} \>
\bigg[\Phi_g^g(z)(1-z) \Big( G(z) + G(1-z)-
 G\Big)\notag\\
&+&n_f\; \Phi_g^q(z)\, \Big(2 Q(z)- G\Big) \bigg].
\label{eq:gpr}
\end{eqnarray}
$\Phi_A^B(z)$ denotes the Dokshitzer-Gribov-Lipatov-Altarelli-Parisi (DGLAP) 
splitting functions \cite{Basics},
\begin{eqnarray*}
\Phi_q^g(z)&\!=&\!C_F\!\!\left(\frac2z+\phi_q^g(z)\right),\,\, 
\Phi_g^g(z)(1-z)\!=\!2N_c\!\!\left(\frac1z+\phi_g^g(z)\right),\\ 
\Phi_g^q(z)&\!=&\!T_R\left[z^2+(1-z)^2\right],
\end{eqnarray*}
where $\phi_q^g(z)=z-2$ and $\phi_g^g(z)=(z-1)(2-z(1-z))$ are regular
as $z\to0$, 
$C_F=(N_c^2-1)/2N_c$, $T_R=n_f/2$ ($N_c=3$ is the number of colors
for $SU(3)_c$ and
$n_f=3$ is the number of light flavors we consider).
The running coupling of QCD ($\alpha_s$) is given by
$$
\alpha_s=\frac{2\pi}{4N_c\beta_0(\ell+y+\lambda)},
\quad \beta_0=\frac1{4N_c}\left(\frac{11}{3}N_c-\frac43T_R\right)
$$ 
and
\begin{equation*}
\ell=\left(1/x\right)\ ,\ y=\ln\left(\kt\big/Q_0\right)\ ,
\ \lambda=\ln\big(Q_0/\lqcd\big),
\end{equation*}
where $Q_0$ is the collinear cut-off parameter. Moreover,
\begin{eqnarray*}
G & \equiv & G(1) = xD_G^h(x,E\Theta_0,Q_0),\cr
Q & \equiv & Q(1) = xD_Q^h(x,E\Theta_0,Q_0).
\end{eqnarray*}
At small $x\ll z$, the fragmentation functions behave as
\begin{equation*}
B(z)\approx \rho_B^{h} \left(\ln\frac{z}{x},\ln\frac{zE\Theta_0}{Q_0}\right)
=\rho_B^h\left(\ln z + \ell, y\right),
\end{equation*}
$\rho_B^{h}$ being a slowly varying function of two logarithmic
variables $\ln (z/x)$ and $y$ that describes the ``hump-backed''
plateau. Since recoil effects should be largest in hard
parton splittings, the strategy followed in this work is
to perform Taylor expansions (first advocated for in \cite{Dremin})
of the non-singular parts of the integrands
in~(\ref{eq:qpr},\ref{eq:gpr}) in powers of
$\ln z$ and $\ln(1-z)$, both considered  small  with respect to
$\ell$ in the hard splitting region  $z\sim 1-z =\cO{1}$
%
\begin{eqnarray}
\label{eq:logic1}
B(z) = B(1) + B_\ell(1) \ln z + \cO{\ln^2z}\ ;\ z\leftrightarrow 1-z\, .
\end{eqnarray}
Each $\ell$-derivative giving an extra  $\sqrt{\alpha_s}$ factor
(see~\cite{RPR2}), the terms $B_\ell(1) \ln z$ and $B_\ell(1)
\ln\left(1-z\right)$ yield NMLLA corrections to the solutions of
(\ref{eq:gpr}).
Making used of (\ref{eq:logic1}), after integrating over 
the regular parts of the DGLAP splitting functions, 
while keeping the singular
terms unchanged, one gets after some algebra
($\gamma_0^2=2N_c \alpha_s/\pi$)~\cite{PAM,PRL}
\begin{eqnarray}\notag
Q(\ell,y)\!&\!=\!&\!\delta(\ell)+\frac{C_F}{N_c}\!\!
\int_0^{\ell}\! \dd \ell'\!\int_0^{y}\! \dd y' \gamma_0^2(\ell'+y')
\Big[1-\tilde a_1\delta(\ell'-\ell)\Big.\\ 
\!&\!+\!&\! \Big.\tilde a_2\delta(\ell'-\ell)
\psi_\ell(\ell',y')  \Big]G(\ell',y'),\label{eq:solq}\\
G(\ell,y)\!&\!=\!&\!\delta(\ell)
+\int_0^{\ell}\! \dd \ell'\!\int_0^{y}\! \dd y' \gamma_0^2(\ell'+y')
\Big[
1 - a_1\delta(\ell'-\ell)\notag\Big.\\
\!&\!+\!&\!\Big.a_2\delta(\ell'-\ell)\psi_\ell(\ell',y')\Big]
G(\ell',y'),
\label{eq:solg}
\end{eqnarray}
with $\psi_\ell(\ell, y)=G_\ell(\ell, y)/G(\ell, y)$. The MLLA coefficients\newline
${\tilde{a}_1=3/4}$ and $a_1\approx0.935$ are computed in~\cite{RPR2} while
at NMLLA, we get
\begin{eqnarray}
\tilde a_2&=&\frac78+\frac{C_F}{N_c}\left(\frac58-\frac{\pi^2}6
\right)\approx0.42,\\
a_2&=&\frac{67}{36}-\frac{\pi^2}6-\frac{13}{18}
\frac{n_fT_R}{N_c}\frac{C_F}{N_c} \approx 0.06\ .
\label{eq:a2}
\end{eqnarray}
Defining $F(\ell,y)=\gamma_0^2(\ell+y)G(\ell,y)$, we can exactly solve 
the self-contained equation (\ref{eq:solg}) by performing the Mellin
transform

\begin{equation}\label{eq:mellin}
F(\ell,y)=\iint\frac{d\omega d\nu}{(2\pi i)^2}e^{\omega\ell}e^{\nu y}
{\cal F}(\omega,\nu).
\end{equation}
Inserting (\ref{eq:mellin}) into (\ref{eq:solg}) we obtain, after some
algebra, the differential equation

\begin{equation}\label{eq:diffeq}
\beta_0\left(\lambda{\cal F}-\frac{\partial{\cal F}}{\partial\omega}-
\frac{\partial{\cal F}}{\partial\nu} \right)=\frac1{\nu}+
\frac{{\cal F}}{\omega\nu}-a_1\frac{{\cal F}}{\nu}-a_2\frac{\omega}{\nu}{\cal F}.
\end{equation}
Finally, after inserting the solution of (\ref{eq:diffeq}) into 
(\ref{eq:mellin}), the new integral representation for the inclusive
spectrum (hump-backed plateau) with NMLLA accuracy reads
\begin{eqnarray}\notag
G(\ell,y)\!&\!=\!&\!(\ell+y+\lambda)\!\!\iint\!\!\frac{d\omega d\nu}{(2\pi i)^2}
e^{\omega\ell}e^{\nu y}\!\int_0^{\infty}\!\!\!\frac{\dd s}{\nu+s}\\
\!&\!\times\!&\!\left(\frac{\omega
\left(\nu+s\right)}{\left(\omega+s\right)\nu}\right)^{\sigma_0}
\left(\frac{\nu}{\nu+s}\right)^
{\sigma_1+\sigma_2}e^{-\sigma_3\,s},
\label{eq:mellinnmlla}
\end{eqnarray}
where 
$$
\sigma_0=\frac{1}{\beta_0(\omega-\nu)},\;
\sigma_1=\frac{a_1}{\beta_0},\;
\sigma_2=\frac{a_2}{\beta_0}(\omega-\nu),\; 
\sigma_3=\frac{a_2}{\beta_0}+\lambda.
$$
However, computing (\ref{eq:mellinnmlla}) numerically is quite a challenging
task. As can be seen, the NMLLA coefficient $a_2$ is very small.
This may explain {\it a posteriori} why the MLLA ``hump-backed plateau''
agrees very well with experimental data~\cite{HBP,DFK}.
Therefore, the NMLLA solution (\ref{eq:mellinnmlla}) of (\ref{eq:solg}) can be approximated
by the MLLA solution of $G(\ell,y)$ ({\it i.e.} taking $a_2\approx0$),
which will be used in the following to compute the inclusive $\kt$-distribution. 
As demonstrated
in \cite{RPR2}, taking the limits $a_2\approx0$ and $\lambda\approx0$ in (\ref{eq:mellinnmlla}), the integral representation can be reduced to the 
known MLLA formula \cite{Basics}
\begin{equation}
G(\ell,y) = 2\ \frac{\Gamma(B)}{\beta_0}\
 \int_0^\frac{\pi}{2}\
  \frac{\dd\tau}{\pi}\,e^{-B\alpha}
\  {\cal F}_B(\tau,y,\ell),
\label{eq:ifD}
\end{equation}
where the integration
is performed with respect to $\tau$ defined by
$\displaystyle \alpha = \frac{1}{2}\ln\frac{y}{\ell}  + i\tau$ and with
\begin{eqnarray*}
{\cal F}_B(\tau,y,\ell) &=& \left[ \frac{\cosh\alpha
-\displaystyle{\frac{y-\ell}{y+\ell}}
\sinh\alpha} 
 {\displaystyle \frac{\ell +
y}{\beta_0}\,\frac{\alpha}{\sinh\alpha}} \right]^{B/2}
\!\!\!I_B(2\sqrt{Z(\tau,y,\ell)}), \cr
&& \cr
&& \cr
 Z(\tau,y,\ell) &=&
\frac{\ell + y}{\beta_0}\,
\frac{\alpha}{\sinh\alpha}\,
 \left(\cosh\alpha
-\frac{y-\ell}{y+\ell}
\sinh\alpha\right),
\label{eq:calFdef}
\end{eqnarray*}
$B=a_1/\beta_0$ and $I_B$ is the modified Bessel function of the first kind.
To get a quantitative idea on the difference between MLLA and NMLLA
gluon inclusive spectrum, the reader is reported
to the appendix B of \cite{PAM}.
The magnitude of $\tilde a_2$, however, indicates that the NMLLA
corrections to the inclusive quark jet spectrum may not be negligible
and should be taken into account. 
After solving (\ref{eq:solg}), the solution of (\ref{eq:solq}) reads
\begin{eqnarray}\label{eq:ratioqg}
Q(\ell,y)\!&\!=\!&\!\frac{C_F}{N_c}\left[G(\ell,y)
+\Big(a_1-\tilde a_1\Big)G_\ell(\ell,y)\right.\\
\!&\!+\!&\!\left.
\left(a_1\Big(a_1-\tilde a_1\Big)+\tilde a_2-a_2\right)G_{\ell\ell}(\ell,y)
\right]+{\cal O}(\gamma_0^2).\notag
\end{eqnarray}
It differs from the MLLA expression given in \cite{PerezMachet} by the term
proportional to $G_{\ell\ell}$, which can be deduced from the subtraction of 
$(C_F/N_c)\times$(\ref{eq:solg}) to Eq.~(\ref{eq:solq}).

\section{Single inclusive $\boldsymbol{k_\perp}-$distribution of charged hadrons in NMLLA}

\begin{figure}[h]
\begin{center}
\includegraphics[height=2.5cm]{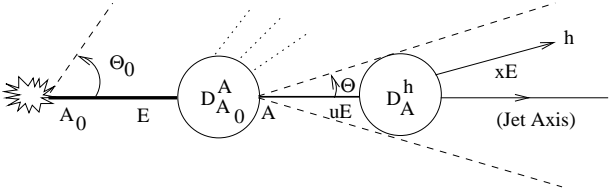}
\caption{\label{fig:distri} Inclusive production of hadron $h$ at angle
$\Theta$ inside a high energy jet of total opening angle
$\Theta_0$ and energy $E$.}
\end{center}
\end{figure}

\begin{figure}
\begin{center}
\includegraphics[height=3.5cm,width=4.2cm]{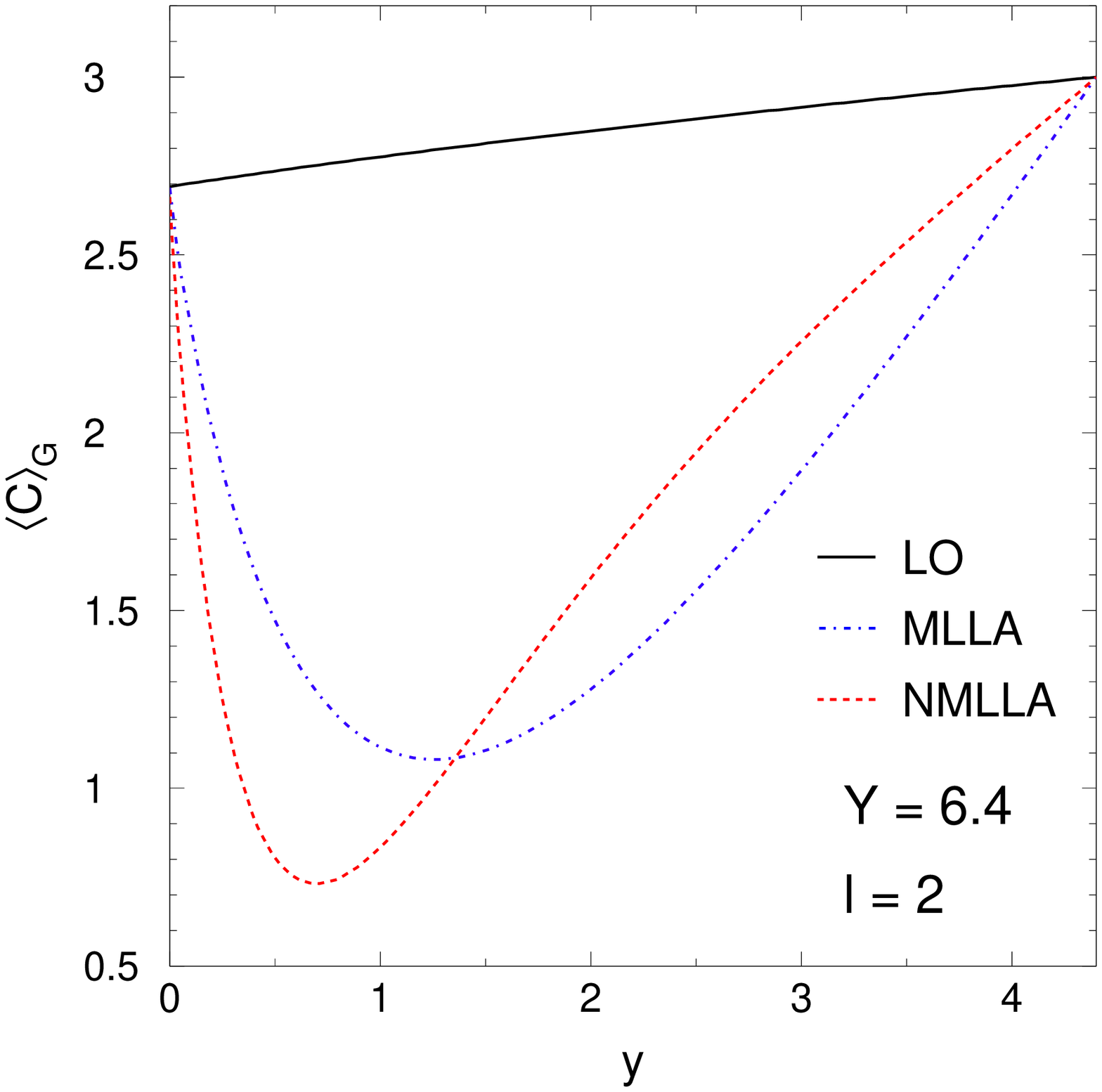}
\includegraphics[height=3.5cm,width=4.2cm]{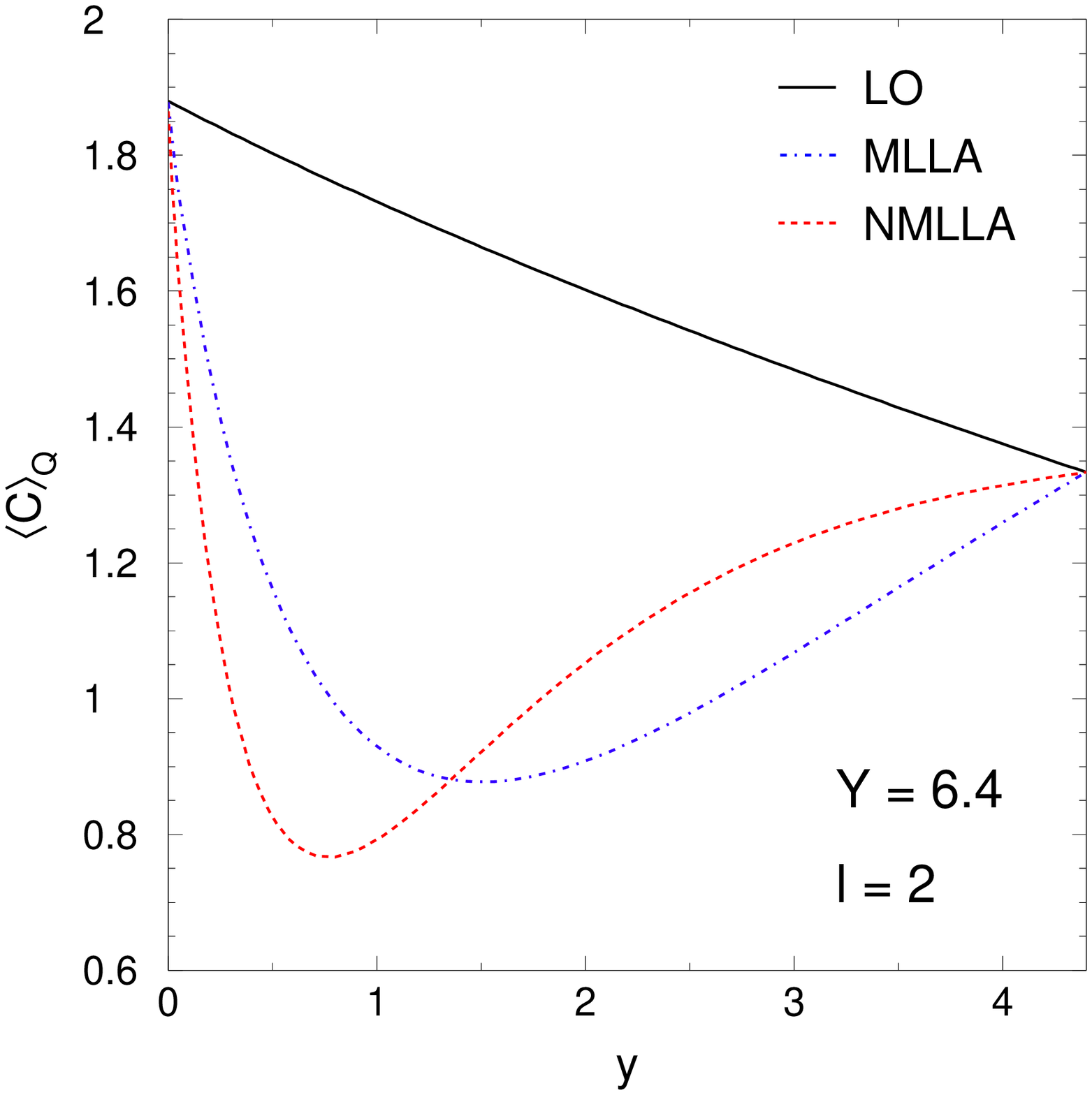}
\caption{\label{fig:CCQ}The color current of a gluon jet (left panel)
and a quark jet (right panel) with $Y_{\Theta_0}
= 6.4$ as a function of $y$ at fixed $\ell=2$ for $\lambda\approx0$.}
\end{center}
\end{figure}

\begin{figure}
\begin{center}
\includegraphics[height=4cm,width=5cm]{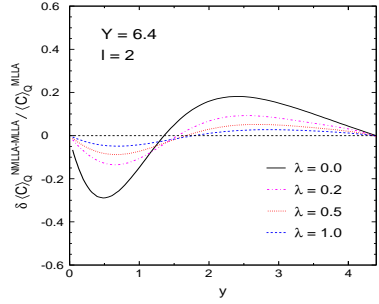}
\caption{\label{fig:CCQlambda}NMLLA corrections to the color current of
a quark jet with $Y_{\Theta_0}=6.4$ and $\ell=2$ for various
values of $\lambda$.}
\end{center}
\end{figure}

\begin{figure}
\begin{center}
\includegraphics[height=7.5cm,width=7.5cm]{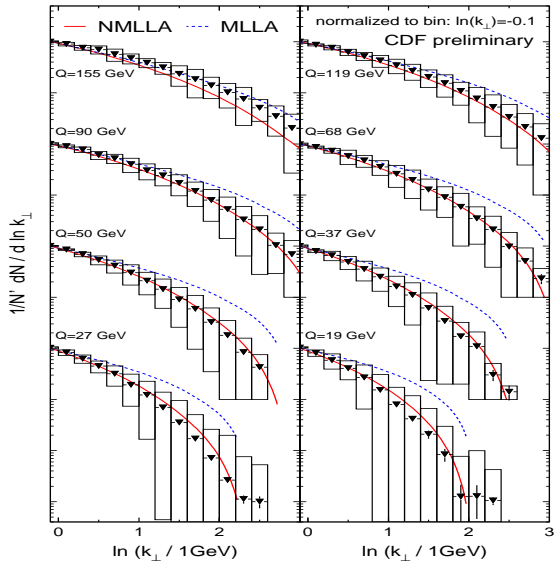}
\caption{\label{fig:CDF-NMLLA}CDF preliminary results for the inclusive
$\kt$ distribution at various hardness $Q$ in comparison to MLLA and
NMLLA predictions at the limiting spectrum ($Q_0=\Lambda_{QCD}$); the boxes are
the systematic errors.}
\end{center}
\end{figure}

\begin{figure}
\begin{center}
\includegraphics[height=4cm, width=5cm]{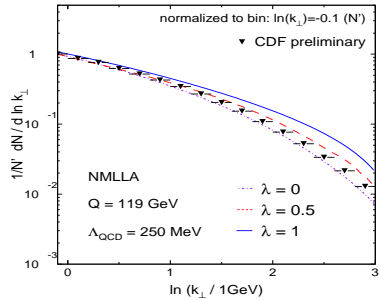}
\caption{\label{fig:CDF-NMLLAlambda}CDF preliminary results ($Q=119$~GeV)
for inclusive $\kt$
distribution compared with NMLLA predictions beyond the limiting spectrum.}
\end{center}
\end{figure}

Computing the single inclusive $k_\perp-$ distribution requires
the definition of the jet axis. The starting 
point of our approach consists in considering
the correlation between two particles
(h1) and (h2) of energies $E_1$ and $E_2$ which form a
relative angle $\Theta$ inside one jet of total 
opening angle $\Theta_0>\Theta$ \cite{DDT}. 
Weighting over the energy $E_2$ of particle (h2), 
this relation leads to the correlation
between the particle (h=h1)
and the energy flux, which we identify
with the jet axis (see Fig.~\ref{fig:distri}) \cite{PerezMachet}. Thus,
the correlation and the relative transverse momentum 
$k_\perp$ between (h1) and (h2) are replaced by the correlation,
and transverse momentum of (h1) with
respect to the direction of the energy flux. 
Finally, we obtain
the double differential
spectrum $\dd^2N/\dd{x}\,\dd\Theta$ of a hadron produced  with energy
$E_1=xE$ at angle
$\Theta$ (or $k_\perp\approx xE\Theta$) with respect to the jet axis.
As demonstrated in~\cite{PerezMachet}, the correlation reads
\begin{equation}
\frac{\dd^2N}{\dd x\,\dd\ln{\Theta}}=
\frac{\dd}{\dd\ln\Theta}F_{A_0}^{h}\left(x,\Theta,E,\Theta_0\right),
\label{eq:DD}
\end{equation}
where $F_{A_0}^{h}$ is given by the convolution
of two fragmentation functions
\begin{equation}
F_{A_0}^{h} \equiv
\sum_{A=g,q}\int_x^1 \dd u D_{A_0}^A\left(u,E\Theta_0,uE\Theta\right)
D_A^{h}\left(\frac{x} {u},uE\Theta,Q_0\right),
\label{eq:F}
\end{equation}
$u$ being the energy fraction of the intermediate parton $A$.
$D_{A_0}^A$ describes
the probability to emit $A$ with energy $uE$ off the parton $A_0$
(which initiates the jet),  taking into account the evolution
of the jet between $\Theta_0$ and $\Theta$. $D_{A}^h$ describes the
probability to produce the hadron $h$ off $A$ with energy fraction $x/u$ and
transverse momentum $\kt\approx uE\Theta\geq Q_0$
(see Fig.~\ref{fig:distri}).
As discussed in \cite{PerezMachet}, the convolution
(\ref{eq:F}) is dominated by $u\sim 1$ and therefore
$D_{A_0}^A\left(u,E\Theta_0,uE\Theta\right)$ is determined by the DGLAP evolution
\cite{Basics}. On the contrary, the distribution
$D_A^{h}\left(\frac{x}{u},uE\Theta,Q_0\right)$ 
at low $x\ll u$ reduces to the hump-backed
plateau,
\begin{equation}
\tilde D_A^h(\ell + \ln u,y) \stackrel{x\ll u}
{\approx} \rho_A^{h}(\ell + \ln u, Y_\Theta + \ln u),
\end{equation}
with $Y_\Theta=\ell+y=\ln E\Theta/Q_0$. Performing the Taylor expansion of
$\tilde D$ to the second order in $(\ln u)$ and plugging it
into Eq.~(\ref{eq:F}) leads to
\begin{eqnarray}
\label{eq:Fdev}
x F_{A_0}^{h}
&\approx& \sum_{A=g,q}\ \int \dd u\, u\, D_{A_0}^A(u,E\Theta_0,uE\Theta)
\tilde D_A^{h}(\ell,y)\\ 
&&\hskip -1cm+\sum_{A=g,q}\ \int \dd u\, u \ln u \, D_{A_0}^A(u,E\Theta_0,uE\Theta)
\frac{\dd \tilde D_A^{h}(\ell,y)}{\dd \ell}\nonumber\\
&&\hskip -1cm+\frac12\sum_{A=g,q} \left[\int \dd u\, u\ln^2 u D_{A_0}^A(u,E\Theta_0,uE\Theta)
\right] \frac{\dd^2 \tilde D_A^{h}(\ell,y)}{\dd\ell^2}.\nonumber
\end{eqnarray}
Indeed, since soft particles are less sensible to the energy balance,
the correlation (\ref{eq:F}) is proved to disappear for these particles, 
leading to the sequence of factorized terms written in (\ref{eq:Fdev}).
The first two terms in Eq.~(\ref{eq:Fdev}) correspond
to the MLLA distribution calculated in \cite{PerezMachet}
when  $\tilde D_A^{h}$ is evaluated at NLO and its derivative at LO.
NMLLA corrections arise from their respective calculation at
NNLO and NLO, and, mainly in practice, from  the third line, which was
computed in \cite{PAM,PRL}.
Indeed, since $x/u$ is small, the inclusive spectrum
$\tilde D_A^h(\ell,y)$ with $A=G,Q$ are given by 
the solutions (\ref{eq:mellinnmlla}) and (\ref{eq:ratioqg})
of the next-to-MLLA evolution equations (\ref{eq:solq})
and (\ref{eq:solg}) respectively. However, because of the smallness
of the coefficient $a_2$ (see (\ref{eq:a2})), $G(\ell,y)$ shows no significant
difference from MLLA to NMLLA. As a consequence, we use the MLLA
expression (\ref{eq:ifD}) for $G(\ell,y)$, and the NMLLA
(\ref{eq:ratioqg}) for $Q(\ell,y)$.
The functions $F_{g}^{h}$ and $F_{q}^{h}$ are related to the gluon
distribution {\it via} the color currents $\langle C\rangle_{g, q}$
defined as:
\begin{equation}
\label{eq:coldef}
x F_{g, q}^{h} = \frac{\langle C\rangle_{g, q}}{N_c}\ G(\ell,y).
\end{equation}
$\langle C\rangle_{g, q}$ can be seen as the average color charge carried
by the parton $A$ due to the DGLAP evolution from $A_0$ to $A$.
Introducing the first and second logarithmic derivatives of
$\tilde D_{A}^{h}$,
\begin{eqnarray*}
\psi_{A,\ell}(\ell,y) &=& \frac{1}{\tilde D_A^h(\ell, y)}
\frac{\dd \tilde D_A(\ell, y)}{\dd\ell}={\cal O}(\sqrt{\alpha_s}),\\
(\psi_{A,\ell}^2+\psi_{A,\ell\ell})(\ell,y)&=&\frac{1}{\tilde D_A^h(\ell, y)}
\frac{\dd^2 \tilde D_A(\ell, y)}{\dd\ell^2}={\cal O}(\alpha_s).
\label{eq:psidef}
\end{eqnarray*}
Eq.~(\ref{eq:Fdev}) can now be written as
\begin{eqnarray*}
x F_{A_0}^{h} \!&\!\approx\!&\! \sum_{A=g,q}\
\Big[ \langle u \rangle_{A_0}^A + \langle u \ln u\rangle_{A_0}^A
\psi_{A,\ell}(\ell,y)\Big.\\
\!&\!+\!&\!\Big. \frac12 \langle u 
\ln^2 u\rangle_{A_0}^A (\psi_{A,\ell}^2+\psi_{A,\ell\ell})(\ell,y) \Big]
\ \tilde D_{A}^h,
\end{eqnarray*}
with the notation
\begin{eqnarray*}
\langle u \ln^i u\rangle_{A_0}^A &\equiv& \int_{0}^{1} \dd u
\ (u\ \ln^i u)\ D_{A_0}^A\left(u,E\Theta_0,uE\Theta\right)\\
&\approx& \int_{0}^{1} \dd u\ (u\ \ln^i u)\ D_{A_0}^A
\left(u,E\Theta_0,E\Theta\right).
\end{eqnarray*}
The scaling violation of the DGLAP fragmentation function
neglected in the last approximation is a NMLLA correction to $<u>$.
It however never exceeds $5\%$ \cite{PAM} of the leading
contribution and is thus neglected in the following.
Using (\ref{eq:coldef}), the MLLA and NMLLA corrections to the leading
color current of the parton $A_0=g,q$ read
\begin{eqnarray}
\delta\langle C \rangle_{A_0}^{\rm MLLA-LO} &=&
N_c\ \langle u \ln u\rangle_{A_0}^g\!\! \psi_{g, \ell}\label{eq:ccmlla}\\
&+&C_{F} \langle u \ln u\rangle_{A_0}^q\!\! \psi_{q, \ell},\notag\\
\delta\langle C \rangle_{A_0}^{\rm NMLLA-MLLA}&=&
N_c\ \langle u \ln^2 u\rangle_{A_0}^g
(\psi^2_{g, \ell}+\psi_{g,\ell\ell})\label{eq:ccnmllabis}\\
&+& C_{F}\ \langle u \ln^2 u\rangle_{A_0}^q\ (\psi^2_{q, \ell}
+ \psi_{q, \ell\ell}).\notag
\end{eqnarray}
The MLLA correction, $\cO{\sqrt{\alpha_s}}$,  was determined
in~\cite{PerezMachet} and the NMLLA contribution, $\cO{\alpha_s}$,
to the average color current is new. The latter can be obtained
from the Mellin moments of the DGLAP fragmentation functions
\begin{equation*}
{\cal D}_{A_0}^A(j,\xi)=\int_0^1 \dd u\,u^{j-1} D_{A_0}^A(u,\xi),
\end{equation*}
leading to
\begin{equation}\label{eq:delta2}
\langle u \ln^2 u\rangle_{A_0}^A = \frac{\dd^2}{\dd j^2}{\cal D}_{A_0}^A
(j,\xi(E\Theta_0)-\xi(E\Theta))\bigg|_{j=2}.
\end{equation}
Plugging~(\ref{eq:delta2}) into the resummed expression $<C>_{g,q}$ of
the color current as written in (\ref{eq:coldef}), these quantities
for gluon and quark jets are determined analytically~\cite{PAM}.
For illustrative purposes, the LO, MLLA, and NMLLA average color
current of a quark jet with 
$Y_{\Theta_0}=6.4$ --~corresponding roughly to Tevatron energies~--
is plotted in Fig.~\ref{fig:CCQ} as a function of $y$, at fixed $\ell=2$
for $\lambda\approx0$.
As discussed in~\cite{PerezMachet}, the MLLA corrections to the LO color
current are found to be large and negative.
As expected, the correction $\cO{\alpha_s}$ from MLLA to NMLLA proves
much smaller; it is negative (positive) at small (large) $y$.
This calculation has also been extended beyond the limiting spectrum,
$\lambda\ne 0$,  to take into account hadronization effects in
the production of ``massive'' hadrons, $m=\cO{Q_0}$~\cite{finitelambda}.
We used, accordingly, the more general 
MLLA solution of (\ref{eq:solg}) with $a_2=0$ 
for an arbitrary $\lambda\ne0$
\begin{eqnarray*}
G\left(\ell,y\right)\!& =&\! \left(\ell+y+\lambda\right)\! \iint\!
\frac{\dd\omega\, \dd\nu}{\left(2\pi i\right)^2}\! e^{\omega\ell+\nu y}\! \int_{0}^{\infty}\frac{\dd s}{\nu+s}\!\\ 
&\times&\left(\frac{\omega
\left(\nu+s\right)}{\left(\omega+s\right)\nu}\right)^{1/\beta_0
\left(\omega-\nu\right)}\left(\frac{\nu}{\nu+s}\right)^
{a_1/\beta_0}e^{-\lambda s},
\end{eqnarray*}
from which an analytic approximated expression was found using the steepest
descent method~\cite{RPR3}; $\sigma_2=0$ and $\sigma_3=\lambda$ have been set
in (\ref{eq:mellinnmlla}). However, $G(\ell, y)$ is
here determined exactly from an equivalent representation in terms of a
single Mellin transform (which reduces to~(\ref{eq:ifD}) as $\lambda\to 0$)~\cite{finitelambda}
\begin{eqnarray}\label{eq:lambdadiff0}
G(\ell,y)&=&\frac{\ell+y+\lambda}{\beta_0\ B\ (B+1)}
\int_{\epsilon-i\infty}^{\epsilon+i\infty}\frac{\dd\omega}{2\pi i}\
e^{\omega\ell}\\
&&\times\  \Phi(-A+B+1, B+2, -\omega(\ell+y+\lambda))\ \ {\cal K}(\omega, \lambda)
\nonumber
\end{eqnarray}
which is better suited for numerical studies. The function ${\cal K}$
appearing in Eq.~(\ref{eq:lambdadiff0}) reads
\begin{equation}
{\cal K}(\omega,\lambda)\ =\ \frac{\Gamma(A)}{\Gamma(B)}\ 
(\omega\ \lambda)^B\ \Psi(A, B+1,\omega\ \lambda),
\end{equation}
where $A=1/(\beta_0\ \omega)$, $B=a_1/\beta_0$, and $\Phi$ and $\Psi$
are the confluent hypergeometric function of the first and second kind,
respectively.
The NMLLA (normalized) corrections to the MLLA result are displayed in
Fig.~\ref{fig:CCQlambda} for different values 
$\lambda=0,0.5,1$, by using (\ref{eq:lambdadiff0}).
It clearly indicates that
the larger $\lambda$, the smaller the NMLLA corrections.
In particular, they  can be as large as $30\%$ at the limiting
spectrum ($\lambda=0$) but no more than $10\%$ for $\lambda=0.5$.
This is not surprising since $\lambda\ne 0$ ($Q_0\ne\lqcd$) reduces
the parton emission in the infrared sector and, thus, higher-order
corrections.
The double differential spectrum\newline $(\dd^2N/\dd\ell\,\dd{y})$,
Eq.~(\ref{eq:DD}), can now be determined from the NMLLA color currents
(\ref{eq:ccnmllabis}) using the MLLA quark and gluon distributions.
Integrating it over $\ell$ leads to the single inclusive $y$-distribution 
(or $\kt$-distribution) of hadrons inside a quark or a gluon jet:
\begin{equation}
  \left(\frac{\dd N}{\dd y}\right)_{g, q}
= \left(\kt\frac{\dd N}{\dd \kt}\right)_{g, q}
= \int_{\ell_{\rm min}}^{Y_{\Theta_0}-y}\;\dd\ell
\;\left(\frac{\dd^2N}{\dd\ell\, \dd y}\right)_{g, q},
\label{eq:lmin}
\end{equation}
where
\begin{equation}
\left(\frac{\dd^2N}{\dd\ell\, \dd y}
\right)_{\tt A_0}\ =\ \frac{1}{N_c}\ \langle C \rangle_{\tt A_0}\ 
\frac{\dd}{\dd{y}}G(\ell,y)\
+\ \frac{1}{N_c}\ G(\ell,y)\ \frac{\dd}{\dd{y}}\langle C \rangle_{\tt A_0}
\label{eq:DDD}
\end{equation}
is the explicit expression of the double differential distribution (\ref{eq:DD}).
The MLLA framework does not specify down to which values of
$\ell$ (up to which values of $x$) the double differential
spectrum $(\dd^2N/\dd\ell\,\dd{y})$
should be integrated over. Since $(\dd^2N/\dd\ell\,\dd{y})$ becomes
negative (non-physical) at small values of $\ell$
(see e.g.~\cite{PerezMachet}), we chose the lower bound $\ell_{\rm min}$
so as to guarantee the positiveness of $(\dd^2N/\dd\ell\,\dd{y})$
over the whole $\ell_{\rm min}\le \ell \le Y_{\Theta_0}$ range
(in practice, $\ell_{\rm min}^g\sim 1$ and $\ell_{\rm min}^q\sim 2$). 
Having successfully computed the single $\kt$-spectra including NMLLA
corrections, we now compare the result with existing data.
The CDF collaboration at the Tevatron
recently reported on preliminary measurements  over a wide
range of jet hardness, $Q=E\Theta_0$, in $p\bar{p}$ collisions at
$\sqrt{s}=1.96$~TeV~\cite{CDF}. CDF data, including systematic errors,
are plotted in
Fig.~\ref{fig:CDF-NMLLA} together with the MLLA predictions of
\cite{PerezMachet} and the present NMLLA calculations, both
at the limiting spectrum ($\lambda=0$) and taking $\lqcd=250$~MeV;
the experimental distributions suffering from large normalization errors,
data and theory are normalized to the same bin, $\ln(\kt/1\,\text{GeV})=-0.1$.
The agreement between the CDF results and the NMLLA distributions over
the whole $\kt$-range is particularly good.
In contrast, the MLLA predictions prove reliable in a much smaller
$\kt$ interval. At fixed jet hardness (and thus $Y_{\Theta_0}$),
NMLLA calculations prove
accordingly trustable in a much larger $x$ interval.
Despite this encouraging agreement with data, the present calculation still
suffers from certain theoretical uncertainties, discussed in detail in
\cite{PAM}. Among them, the variation of $\lqcd$ --giving NMLLA
corrections-- from the default value $\lqcd = 250$ MeV to $150$ MeV and
$400$ MeV affects the normalized $\kt$-distributions by roughly $20\%$
in the largest $\ln(\kt/1~\text{GeV})=3~\text{GeV}$-bin at $Q=100$ GeV. Also, cutting
the integral (\ref{eq:lmin}) at small values of $\ell$ is somewhat arbitrary.
However, we checked that changing $\ell_{min}$ from $1$ to $1.5$ modifies
the NMLLA spectra at large $\kt$ by $\sim 20\%$ only
\footnote{The effect of varying $\ell_{min}$ is more dramatic at MLLA.}. 
%
%
Finally, the kt-distribution is determined with respect to the jet energy
flow (which includes a summation over secondary hadrons in
energy-energy correlations). In experiments, instead, the jet axis
is determined exclusively from all particles inside the jet.
The question of the matching of these two definitions goes beyond
the scope of this letter.
The NMLLA $\kt$-spectrum has also been calculated beyond the limiting
spectrum, by plugging (\ref{eq:lambdadiff0}) into (\ref{eq:lmin}), as 
illustrated in Fig.~\ref{fig:CDF-NMLLAlambda}.
However, the best description of CDF preliminary data is reached at the limiting
spectrum, or at least for small values of $\lambda\lesssim 0.5$, which is not too
surprising since these inclusive measurements 
mostly involve pions. Identifying produced hadrons would offer the
interesting possibility to check a dependence
of the shape of $\kt$-distributions on the hadron species,
such as the one predicted in Fig.~\ref{fig:CDF-NMLLAlambda}.
Moreover, the softening of 
the $\kt$-spectra with increasing hadron masses predicted in 
Fig.~\ref{fig:CDF-NMLLAlambda} 
is an observable worth to be measured, as this would provide an 
additional and independent check of the LPHD hypothesis beyond the limiting spectrum. 
This could only be achieved if the various species of hadrons inside a jet can be identified experimentally. Fortunately, it is likely to be the case at the LHC, where the ALICE~\cite{identificationalice}
and CMS~\cite{identificationcms} experiments at the Large Hadron Collider have good 
identification capabilities at not too large transverse momenta.

\section{Conclusion}

To summarize, single inclusive $\kt$-spectra inside a jet are
determined including higher-order $\cO{\alpha_s}$ (i.e. NMLLA) corrections
from the Taylor expansion of the MLLA evolution equations and beyond the
limiting spectrum, $\lambda\ne 0$. The agreement between NMLLA predictions
and CDF preliminary data in $p\bar{p}$ collisions at the Tevatron is
impressive, indicating very small overall
non-perturbative corrections and giving further support
to LPHD \cite{LPHD}. The MLLA evolution equations 
for inclusive enough variables prove once more~(see e.g.~\cite{Basics})
to include reliable information
at higher orders than MLLA.


\begin{thebibliography}{50}
\bibitem{HBP}
Yu.L. Dokshitzer, V.S. Fadin, V.A. Khoze, Phys. Lett. {\bf B 115} (1982) 242;\\
Ya.I. Azimov, Yu.L. Dokshitzer, V.A. Khoze, S.I. Troian, Z. Phys. {\bf C 31} (1986) 213;\\
C.P. Fong, B.R. Webber, Phys. Lett. {\bf B 229} (1989) 289.

\bibitem{KhozeOchs}
V.A. Khoze, W. Ochs, Int. J. Mod. Phys. {\bf A 12} (1997) 2949.

\bibitem{LPHD}
Ya.I. Azimov, Yu.L. Dokshitzer,
Z. Phys {\bf C 27} (1985) 65;\\
Yu.L. Dokshitzer, V.A. Khoze, S.I. Troian, J. Phys. {\bf G 17} (1991) 1585.

\bibitem{PerezMachet}
R. P\'erez Ramos, B. Machet,
JHEP {\bf 04} (2006) 043.

\bibitem{Basics}
Yu.L. Dokshitzer, V.A. Khoze, A.H. Mueller, S.I. Troyan, 
(Editions Fronti\`eres, Gif-sur-Yvette, 1991).

\bibitem{RPR2}
R. P\'erez Ramos, JHEP {\bf 06} (2006) 019 and references therein.

\bibitem{DokKNO}
Yu.L. Dokshitzer,
Phys. Lett. {\bf B 305} (1993) 295.

\bibitem{CuypersTesima}
F. Cuypers, K. Tesima,
Z. Phys. {\bf C 54} (1992) 87.

\bibitem{PAM} R. Perez-Ramos, F. Arl\'eo \& B. Machet,
Phys. Rev. D {\bf 78} (2008) 014019.

\bibitem{KLO}
V.A. Khoze, S. Lupia, W. Ochs, Phys. Lett. {\bf B 386} (1996) 451.

\bibitem{lepdelphi}
DELPHI Collaboration: J. Abdallah {\it et al.},
Phys. Lett. {\bf B 605} (2005) 37.

\bibitem{PRL}
F. Arleo, R. P\'erez-Ramos,  B. Machet, Phys. Rev. Lett. {\bf 100} (2008) 052002.

\bibitem{CDF}
S. Jindariani,  A. Korytov \& A. Pronko,
CDF report CDF/ANAL/JET/PUBLIC/8406 (March 2007),\hfill\break
www-cdf.fnal.gov/physics/new/qcd/ktdistributions\hfill\break\_06/cdf8406\_Kt\_jets\_public.ps.

\bibitem{DFK} Yu.L. Dokshitzer, V.S. Fadin, V.A. Khoze, Z. Phys {\bf C 18} (1983) 37.

\bibitem{Dremin}
I.M. Dremin, Phys. Usp. {\bf 37} (1994) 715;\\
{\it ibid.}, Usp. Fiz. Nauk {\bf 164}
(1994) 785;\\
I.M. Dremin, J.W. Gary, Phys. Rep. {\bf 349}
 (2001) 301;\\
I.M. Dremin, V.A. Nechitailo, Mod. Phys. Lett. {\bf A9} (1994) 1471.

\bibitem{DDT}
Yu.L. Dokshitzer, D.I. Dyakonov \& S.I. Troyan, Phys. Rep. {\bf 58} (1980) 270.

\bibitem{RPR3}
R. P\'erez Ramos, JHEP {\bf 09} (2006) 014.

\bibitem{finitelambda}
Yu.L. Dokshitzer, V.A. Khoze, S.I. Troian, Z. Phys. {\bf C 55} (1992) 107;\\
Yu.L. Dokshitzer, V.A. Khoze, S.I. Troian, Int. J. Mod. Phys. {\bf A 7} (1992) 1875;\\
V.A. Khoze, W. Ochs, J. Wosieck, and references therein.

\bibitem{identificationalice}
ALICE collaboration, J. Phys. {\bf G32} (2006) 1295.

\bibitem{identificationcms}
CMS collaboration, D. d'Enterria (Ed.) {\it et al.}, J. Phys. {\bf G34} (2007) 2307.


\end{thebibliography}
\end{document}